\documentclass[10pt,journal,compsoc]{IEEEtran}
\ifCLASSOPTIONcompsoc
  \usepackage[nocompress]{cite}
\else
  \usepackage{cite}
\fi

\usepackage{cite}
\usepackage{amsmath,amssymb,amsfonts,amsthm}
\usepackage{graphicx}
\usepackage{textcomp}

\usepackage{booktabs, multirow, threeparttable, subfigure}
\usepackage{algorithm, algorithmic}
\usepackage{xspace}
\usepackage{hyperref}

\newtheorem{definition}{Definition}

\DeclareMathOperator*{\argmin}{arg\,min}

\makeatletter
\DeclareRobustCommand\onedot{\futurelet\@let@token\@onedot}
\def\@onedot{\ifx\@let@token.\else.\null\fi\xspace}
\def\eg{\emph{e.g}\onedot} \def\Eg{\emph{E.g}\onedot}
\def\ie{\emph{i.e}\onedot}

\def\etal{\emph{et al}\onedot}

\def\UrlAlphabet{
      \do\a\do\b\do\c\do\d\do\e\do\f\do\g\do\h\do\i\do\j
      \do\k\do\l\do\m\do\n\do\o\do\p\do\q\do\r\do\s\do\t
      \do\u\do\v\do\w\do\x\do\y\do\z\do\A\do\B\do\C\do\D
      \do\E\do\F\do\G\do\H\do\I\do\J\do\K\do\L\do\M\do\N
      \do\O\do\P\do\Q\do\R\do\S\do\T\do\U\do\V\do\W\do\X
      \do\Y\do\Z}
\def\UrlDigits{\do\1\do\2\do\3\do\4\do\5\do\6\do\7\do\8\do\9\do\0}
\g@addto@macro{\UrlBreaks}{\UrlOrds}
\g@addto@macro{\UrlBreaks}{\UrlAlphabet}
\g@addto@macro{\UrlBreaks}{\UrlDigits}

\begin{document}

\title{Streaming Local Community Detection through Approximate Conductance}

\author{
    Yanhao~Yang,
    Meng~Wang,
    David~Bindel,
    Kun~He%,~\IEEEmembership{Senior~Member,~IEEE}% <-this % stops a space
    \IEEEcompsocitemizethanks{
        \IEEEcompsocthanksitem 
        The first two authors 
        %Y. Yang and M. Wang are
        contributed equally to this work.
        \IEEEcompsocthanksitem The first two authors and Kun He are with the School of Computer Science and Technology, Huazhong University of Science and Technology, Wuhan, China, 430074. David Bindel is with the Department of Computer Science, Cornell University, Ithaca, NY 14853-5169. %bindel@cornell.edu
        Kun He is the corresponding author. \protect\\
        E-mail: brooklet60@hust.edu.cn.}% 
    % \thanks{Manuscript received September 30, 2021; revised dd/mm/yyyy.}
}

\IEEEtitleabstractindextext{%
\begin{abstract}
% Community is a universal structure in various complex networks, and community detection is a fundamental task for network analysis. With the rapid growth of network scale in the era of big data, networks are massive, raising the need from global community detection to local community mining, and in many real-world scenarios networks are changing rapidly and naturally that could be modeled as graph streams. 
Community is a universal structure in various complex networks, and community detection is a fundamental task for network analysis. With the rapid growth of network scale, 
%in the era of big data, 
networks are massive, changing rapidly and could naturally be modeled as graph streams. 
Due to the limited memory and access constraint in graph streams, existing non-streaming community detection methods are no longer applicable. This raises an emerging need for online approaches.
In this work, we consider the problem of uncovering the local community containing a few query nodes in graph streams, termed \textit{streaming local community detection}. 
This is a new problem raised recently that is more challenging for community detection and only a few works address this online setting. 
Correspondingly, we design an online single-pass streaming local community detection approach. Inspired by the ``local'' property of communities, our method samples the local structure around the query nodes in graph streams, and extracts the target community on the sampled subgraph using our proposed metric called the \textit{approximate conductance}.  
Comprehensive experiments show that our method remarkably outperforms the streaming baseline on both effectiveness and efficiency, and even achieves similar accuracy comparing to the state-of-the-art non-streaming local community detection methods that use static and complete graphs. 
% We also provide further experiments to show the proposed method is scalable to the network scale and density, and exhibits robustness on the detection accuracy for various community sizes and densities.  
\end{abstract}

}

\maketitle

\IEEEdisplaynontitleabstractindextext

% \fi
\IEEEpeerreviewmaketitle

\section{Introduction}
% please using data streaming model and streaming model rather than data stream model and stream model
%  first paragraph: community detection
% motivation of local community detection: which is more suitable for some settings than global community detection.
%Networks~\footnote{In this paper, we interchangeably use terms \textit{network} and \textit{graph}.} 
\IEEEPARstart{N}{etworks} exhibit natural structures representing entities and their relationships for complex systems in various domains, \eg, society, biology, communication, and Wide World Web, in which vast amounts of data are constantly being generated.
As a fundamental task of network analysis, community detection aims to uncover groups of densely connected nodes, termed communities, and has attracted extensive attention. 
Much effort has been devoted to the global community detection for decades~\cite{newman2004finding, kannan2004clusterings, blondel2008fast, lancichinetti2011finding, coscia2012demon}. 
In recent years, there has been a growing interest in exploring the local community containing a few query nodes~\cite{andersen2006local, cui2013online, huang2014querying, kloster2014heat,he2015detecting, li2015uncovering, veldt2016simple, luo2020local}. 
%yuan2017index, li2018local, he2019krylov}. 
From the perspective of computational cost, the problem of local community detection is more suitable for uncovering the community structure for nodes of interest on large-scale networks, considering the real-world scenarios like recommendation systems and political activism study~\cite{weber2013secular}. 

% second paragraph: real-world scenarios, massive and naturally streaming
% motivation of streaming community detection: massive network, some networks are naturally streaming (when the online social network generating and some information can be observed once and change faster.)
Existing community detection methods, either globally or locally, focus on mining \textit{static} and \textit{complete} networks, which takes a considerable overhead on the memory. 
With the rapid growth of the network scale, the memory space required to handle such massive networks becomes unacceptable for practical applications. 
\Eg, in 2020, Twitter has 340 million users, and each user has an average of 707 followers; Facebook has 1.79 billion daily active users, and each user has an average of 338 friends.
When these websites are constructed as networks, we will obtain two massive networks with millions to billions of nodes and billions to trillions of edges, which occupy petabytes of storage space if adjacency matrices represent them. It is unrealistic to load such networks entirely into the main memory, even in distributed systems~\cite{liakos2017realizing}.
Moreover, under the real-world scenarios, networks are usually changing rapidly, and relationships are only established instantaneously, \eg, in email exchange network or product co-purchase network. Such networks generated by online transient interactions are naturally streaming data~\cite{ahmed2020adaptive} and can be modeled as \textit{graph streams}.

% challenge of graph stream: limited work memory, at most O(ploy|V|); can only access the data in it arrives order; naturally streaming network, the global information is unknown before process the stream.
% challenge of streaming community detection: compare to triangle sampling for graph stream, the community is a higher order and more complex structure.
We consider the problem of local community detection in graph streams, termed \textit{streaming local community detection} (see Definition \ref{def:streaming}). 
Graph stream is a streaming model in which the graph data, \eg, edges of the graph, are arranged as a stream based on their generation order. 
The stream is sometimes infinite and the newly generated data will be added in the end. 
Given a massive and naturally streaming network $G=(V,E)$, the graph stream can organize the data with limited memory. 
There is another side to this coin, data in the stream can only be accessed \textit{once} in the order it arrives, and methods for graph stream mining should only occupy at most $O(|V|P(\log|V|) )$ memory \cite{mcgregor2014graph}, where $P(\cdot)$ indicates a polynomial function. % In other words, the detection is on-the-fly using partial knowledge, and the global information about the network is \textit{unknown} before the stream is processed. 
Furthermore, the global information about the network is \textit{unknown} before the stream is completely processed. 
%Furthermore, for naturally streaming networks, since entities and interactions can only be observed when they occur, the corresponding data in the stream are only accessible \textit{once} and the global information about the network is \textit{unknown} before the stream is processed. 

% streaming community detection, and existing methods that work on static and complete networks, cannot be applied in the streaming model. There is not enough main memory to cache all data in the graph streams.
% As the above discussion, graph stream is naturally representation of real-world networks while it also has many restrictions on the methods in it. In such a model, uncovering community, which is a more complex structure than motifs, is crucial and challenging. 
Recently, there have been several works in graph streams.
Researchers mainly focus on mining relatively simple structures in graph streams, such as counting triangles~\cite{bar2002reductions,ahmed2017sampling} and wedges~\cite{ahmed2017sampling}, sampling motifs \cite{ahmed2020adaptive} and densest subgraphs~\cite{bahmani2012densest}. 
Compared to these structures, the community is much more complex, and hence it is more challenging to uncover community in graph streams.
And for existing local community detection methods working on static and complete networks~\cite{huang2014querying, kloster2014heat, veldt2016simple, he2019krylov}, %li2018local, he2019krylov}, % yuan2017index
%some constraints of the streams are so restrictive that existing local community detection methods are no longer applicable. 
the limited memory and access constraint in graph streams make these methods no longer applicable. 
%It is crucial to design streaming local community detection methods.% that can be fully adapted to graph streams.

In this work, we propose an \textit{online single-pass} streaming local community detection method, called SCDAC (Streaming Community Detection via Approximate Conductance), for uncovering the target local community containing the query nodes in graph streams. Inspired by the ``local'' property of communities \cite{conte2018d2k}, we first sample a subgraph covering the local community structure around the query nodes in the arriving graph stream. Then following the spirit of \textit{seed set expansion}~\cite{andersen2006local, kloster2014heat, li2015uncovering, he2019krylov}, we regard the query nodes as a seed-set and expand it to a comparatively large node set that exhibits high community quality on the sampled subgraph. 

However, the entire graph cannot be cached in graph streams, and many metrics of community quality are unavailable, like conductance~\cite{kannan2004clusterings} and modularity~\cite{newman2006modularity}. Moreover, the subgraph sampled in the graph streams is usually incomplete for the exact local structure around the query nodes, and the local metric adopted by existing local community detection methods on complete subgraph~\cite{li2015uncovering, he2019krylov} are inaccurate on such subgraphs. To address this issue, we propose a metric called \textit{approximate conductance} to measure the community quality in such a scenario. Specifically, the new metric aims to approximate the exact conductance of the community on the incomplete subgraph sampled in graph streams. 

The computational cost of our method, including the time and space complexity, is thoroughly analyzed (in Section \ref{sec:complexity}). And the effectiveness and efficiency of our method are also verified by extensive experimental results.

% contributions: we formulate streaming local community detection; we propose a \textit{online} streaming local community detection method; for graph stream, we propose a new metric measure community quality, termed as approximate conductance.
Our main contributions are summarized as follows:
\begin{itemize}
%    \item We formulate the \textit{streaming local community detection} problem, in which a graph is specified as a stream consisting of all edges.
    \item We design an \textit{online single-pass} streaming local community detection method, which uncovers the target local community for query nodes in graph streams without any prior knowledge. 
    \item We propose a new metric called \textit{approximate conductance} to measure the community quality on the subgraph sampled in graph streams.
    % \item Inspired by the idea of path compression in union-find set, we maintain the intercepted nodes in a distance tree to speedup the calculation.% on the distance of a node to the query node set.
    \item Extensive experiments on real-world networks from various domains demonstrate the superiority of our method in terms of effectiveness and efficiency.
\end{itemize}

\label{sec:introduction}

\section{Background}
% In this section, we present some work and concepts related to this paper, as well as clarify the differences with our work.
\subsection{Problem Formulation}
\label{sec:formulation}
Consider a network modeled as an undirected and unweighted graph $G = (V, E)$, where $V = \{v_i|i=1, 2, ...\}$ represents the node set and $E=\{e_i|i=1, 2, ...\}$ represents the edge set. Given a subset of nodes $T$ called the query nodes (or query node set) contained in a target community $C^*$. Generally, $|T| \ll |C^*| \ll |V|$. The task of the local community detection is to uncover the target community $C^*$ containing the query node set $T$ on the static complete graph $G$. 

% streaming local community detection
In view of the massive and naturally streaming networks, we introduce streaming local community detection as follows. 
\begin{definition} [Streaming local community detection]
\label{def:streaming}
Consider a stream $\mathcal{S}$ as an arbitrarily ordered sequence of all the edges in network $G = (V,E)$, denoted as $\langle e_1, e_2, ..., e_i, ... \rangle$. Given a query node set $T$, the task of streaming local community detection is to uncover the target community $C^*$ containing the query node set $T$ using the stream $\mathcal{S}$ under the limit of $O(|V|P(\log|V|))$ work memory, where $P(\cdot)$ is a polynomial function. 
\end{definition}

The above definition is a general description of the streaming local community detection.
In this work, we consider a more demanding streaming model, where the edges in the graph stream are only available once, rather than a sliding-window model~\cite{mcgregor2014graph}. In more detail, when processing an edge $e_i$ in stream $\mathcal{S}$, the past edges cannot be accessed again, and the subsequent edges in the stream are unknown, \ie, $\forall j \neq i$, $e_j$ is not accessible when $e_i$ arrives.

\begin{figure*}[!t]
    \centering
    \includegraphics[width=\linewidth]{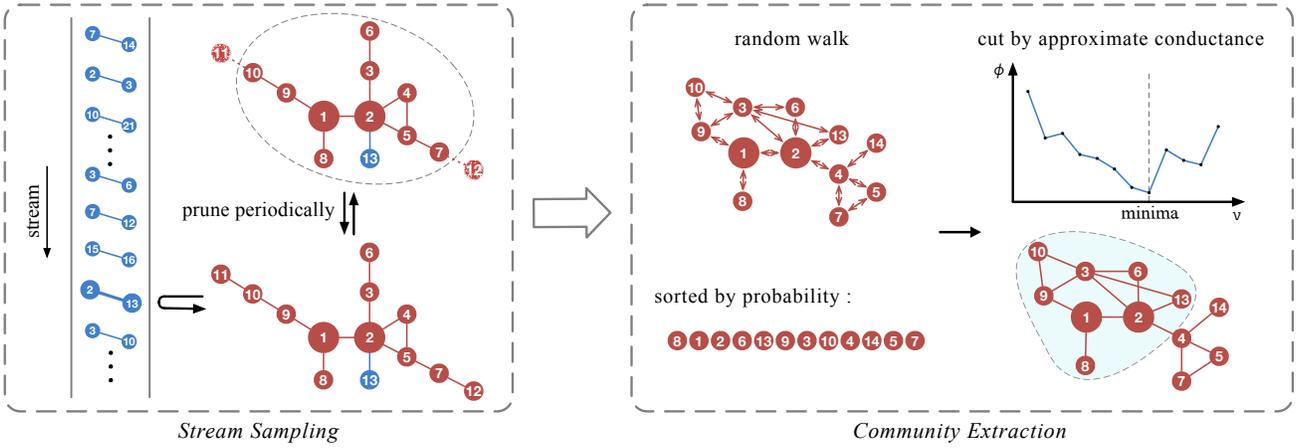}
    \caption{An overview of the proposed SCDAC method. The query nodes are in larger dark red.}
    \label{fig:overview}
\end{figure*}

\subsection{Related Work}
\subsubsection{Local community detection}% on static network}
In contrast to the global community detection~\cite{newman2004finding, kannan2004clusterings, blondel2008fast, lancichinetti2011finding, coscia2012demon, li2019edmot, ye2019discrete, li2020community}, which aims to find all communities in a network, the local community detection~\cite{veldt2016simple, cui2013online, yuan2017index, huang2014querying, andersen2006local, kloster2014heat, li2015uncovering, bian2018multi, luo2020local, bian2020rethinking} is a query-oriented problem to find the target community containing the query node(s).

The phase of community extraction in our method falls into the category of seed-set expansion, a popular category of methods for local community detection~\cite{andersen2006local, kloster2014heat, li2015uncovering, he2019krylov}. The general idea of seed-set expansion is to expand the seed-set along the network topology and obtain a community by optimizing a scoring function. 
Common diffusion methods include random walk~\cite{andersen2006local}, heat kernel~\cite{kloster2014heat} and local spectral approximation~\cite{li2015uncovering, he2019krylov}. Metrics measuring the community quality are usually adopted as the scoring function, such as modularity~\cite{newman2006modularity} and conductance~\cite{kannan2004clusterings}.

Especially, PRN~\cite{andersen2006local} is a pioneer work of seed-set expansion, which adopts lazy random walk as the diffusion method and uncovers the community with local minimum conductance. Kloster and Gleich~\cite{kloster2014heat} adopt heat kernel as the expansion method and present a deterministic local algorithm to compute the heat kernel diffusion by coordinate relaxation. LEMON~\cite{li2015uncovering} and LOSP~\cite{he2019krylov} perform short random walks and span an approximate invariant subspace termed the local spectral subspace, and obtain the membership vector in the subspace. 

All the methods discussed above run on static and complete graphs, which are required to be loaded into the main memory entirely. This limits the application of local community detection on massive and streaming networks.

\subsubsection{Streaming community detection}
As a feasible approach handling massive networks, graph stream has become increasingly popular 
in recent years~\cite{mcgregor2014graph}. A considerable amount of literature on graph stream model has been proposed, covering many areas of network and graph analysis, including estimating connectivity~\cite{feigenbaum2005graph,ahn2012analyzing}, sampling subgraph~\cite{bar2002reductions,bahmani2012densest,ahmed2017sampling,ahmed2020adaptive}, outlier detection~\cite{aggarwal2011outlier}, link prediction~\cite{zhao2016link}, and maintaining the projection of graphs~\cite{tang2016graph,khan2016query,gou2019fast}.

The problem of community detection in graph stream (also called streaming community detection) has also attracted certain attention. Yun \etal~\cite{yun2014streaming} propose a method to detect communities in a stream composed of the columns of the adjacency matrix. Hollocou \etal~\cite{hollocou2017linear} consider a stream consisting of edges and propose a streaming community detection method based on the assumption that an edge randomly selected in the stream is more likely to be inside the community. However, both of them focus on detecting all communities of a network and belong to the category of streaming global community detection. While very little work has studied streaming local community detection. To our knowledge, there is only one work of Liakos \etal~\cite{liakos2020rapid} that aims to detect a local community containing the query nodes in graph streams, and they propose a method called CoEuS. 

Our goal belongs to the same category as CoEuS, but the idea and method are distinct. CoEuS greedily expands the community in the streaming model, while we seek an optimal community on the subgraph intercepted from the streaming model. Extensive experiments show that our method significantly outperforms CoEuS in terms of effectiveness and efficiency.

\section{Algorithm}
In this section, we introduce the proposed algorithm in detail. 
As nodes in the same community are densely connected, the distance between each pair of nodes inside a community is usually small. Under the local community detection scenario, the target community $C^*$ must lurk in the neighborhood of the query nodes. Therefore, we first sample the local structure around the query node set $T$ in graph stream $\mathcal{S}$, then extract a high-quality community containing these query nodes on the local structure. Fig. \ref{fig:overview} provides an overview of our method.

% \begin{figure*}[!t]
%     \centering
%     \includegraphics[width=\linewidth]{figures/overview.eps}
%     \caption{An overview of the proposed SCDAC method. The query nodes are in larger dark red.}
%     \label{fig:overview}
% \end{figure*}

\subsection{Stream Sampling}
\label{sec:sampling}
We first provide several definitions on undirected and unweighted graphs. The distance from node $u$ to node $v$, termed $dist(u,v)$, is the length of the shortest path between the two nodes. The distance from node $u$ to a node set $T$, termed $dist(u,T)$, is the minimum distance from $u$ to any node in $T$, \ie, $dist(u, T) = \min_{\forall v \in T} dist(u,v)$. 
The $k$-hop neighborhood of node $v$ is a set of nodes whose distance to $v$ is no more than $k$, i.e. $\{ u | dist(u,v) \leq k \}$. Similarly, the $k$-hop neighborhood of a node set $T$ is a set of nodes whose distance to $T$ is no more than $k$, i.e. $\{ u | dist(u,T) \leq k \}$.

% This part needs to describe the challenges of sample the local structure (or subgraph), and how we do.
The \textit{local structure} around a query node set $T$, defined as the subgraph induced by nodes in $T$ and their $k$-hop neighbors ($k$ is a small positive integer)~\cite{conte2018d2k}. The goal of the first phase is to sample such a local structure for the query node set $T$. 
However, as we can only access each edge in the stream once and have no prior information on the coming edges before they arrive, it is tough to sample this local structure precisely, even for $k = 1$. And the greater the value of $k$ is, the less accurate the sampled subgraph is.
Thus, the goal of this phase is to sample subgraph $G_s = (V_s, E_s)$, which can approximately cover the $k$-hop neighborhood of the query nodes $T$.

Also, the subgraph size, denoted as $|V_s|$, is related to the power of node degrees. The sampled subgraph will be expanded too large in some networks with nodes in relatively large average degrees. \textit{E.g.}, in Youtube network (in Table \ref{tab:datasets}), a sampled subgraph covering the 4-hop neighborhood of a few query nodes could reach millions of nodes, but we are only interested in nodes densely connected around the query nodes. 
To address this issue, we only keep a certain number of nodes closest to the query nodes in the sampled subgraph. Furthermore, we consider that some nodes are actually close to the query nodes while far away from $T$ when they first join the subgraph. Then we give these nodes a ``probation period'' to not be expelled immediately, \ie, we prune the subgraph to $ps$ nodes only when every $pc$ streaming edges are processed. And $pc$ and $ps$ are termed the pruning cycle and pruning size, respectively.

Details of the stream sampling are presented in Algorithm \ref{alg:sampling}. In the end, we output the sampled subgraph $G_s$ and an array $D$, recording the accumulated degree of all nodes in the stream. 

\renewcommand{\algorithmicrequire}{\textbf{Input:}}
\renewcommand{\algorithmicensure}{\textbf{Output:}}
\begin{algorithm}
\begin{algorithmic}[1]
\caption{Stream sampling}%Sample local structure in graph stream}
\label{alg:sampling}
\REQUIRE Graph stream $\mathcal{S}$, query nodes $T$.
\ENSURE Sampled subgraph $G_s = (V_s, E_s)$, degree array $D$.%$\mathbf{D}$.
\renewcommand{\algorithmicrequire}{\textbf{Parameters:}}
\REQUIRE Number of hops $k$, pruning cycle $pc$, pruning size $ps$.
\STATE $D \leftarrow \mathbf{0}$
\STATE $G_s \leftarrow (T, \emptyset)$
\FOR{\textbf{each} coming $e_i = (u,v) \in \mathcal{S}$}
    \STATE $D[u] \leftarrow D[u] + 1$
    \STATE $D[v] \leftarrow D[v] + 1$
    % consider simplify below commands
    \IF{$dist(u,T) \leq k$ \AND $dist(v,T) \leq k$ in $G_s'=(V_s, E_s \cup \{(u,v)\})$}
        \STATE $V_s \leftarrow V_s \cup \{u, v\}$
        \STATE $E_s \leftarrow E_s \cup \{(u,v)\}$
    \ENDIF
    \IF{$i | pc$ = 0}
        \STATE sort $V_s$ in ascending order by $dist(v, T)$.
        \STATE $V_s \leftarrow \{v_j | v_j \in V_s \land j \leq ps \}$
        \STATE $E_s \leftarrow \{(u,v) | u \in V_s \lor v \in V_s \}$
    \ENDIF
\ENDFOR
\end{algorithmic}
\end{algorithm}

\begin{figure}
    \centering
    \includegraphics[width=\linewidth]{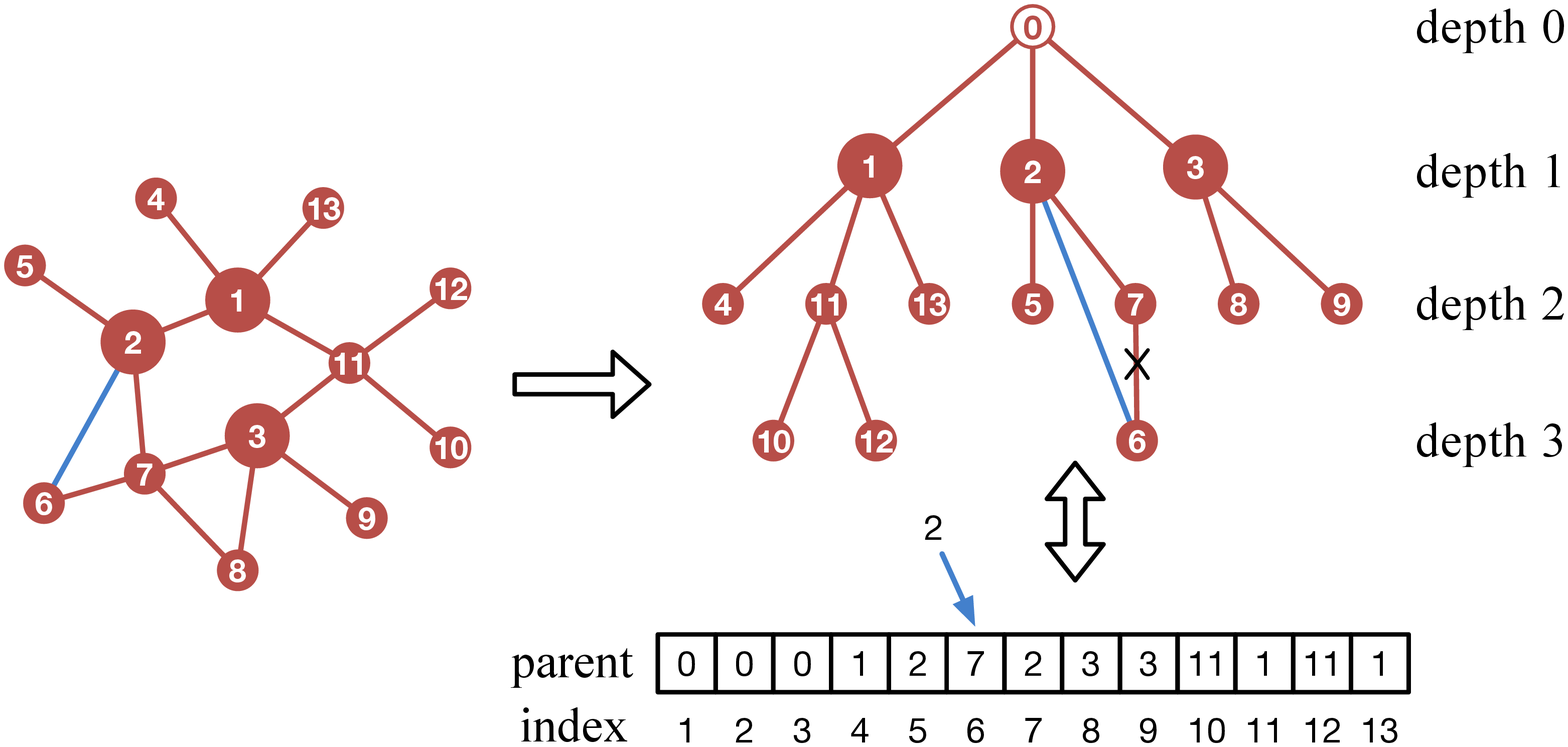}
    \caption{An example of organizing the node set of the sampled subgraph in the distance tree, where \{1,2,3\} are the query nodes, and edge (2,6) is the coming edge of the graph stream.
    }
    \label{fig:tree}
\end{figure}

In Algorithm \ref{alg:sampling}, we need to calculate $dist(v, T)$ many times for a node $v$ and a node set $T$. 
This is a typical single source shortest path problem in the graph, and the well-known solution is based on Breath-First Search in the time complexity of $O(|V_s| + |E_s|)$. Inspired by the idea of path compression in union-find set, we consider to maintain node set $V_s$ in a distance tree (see example on the right of Fig. \ref{fig:tree}) to speedup the calculation on $dist(\cdot,T)$. Node $0$ is a dummy root node, the query nodes in $T$ are with depth $1$, and other nodes are the children of their adjacent node which are in their shortest paths to the query nodes, \ie, for node $v$,
\begin{equation}
    parent(v) = \argmin_{\forall u \in N(v)}{dist(u,T)},
\end{equation}
where $N(v)$ denotes the set of nodes adjacent to $v$ in the sampled subgraph $G_s$. In this way, the problem of calculating the distance from node $v$ to the query node set $T$ can be converted into the problem of calculating the depth of node $u$ in the tree. Formally, we have:
\begin{equation}
    dist(v, T) = 
    \begin{cases}
        depth(v) - 1 & \text{if} \quad v \in V_s, \\
        \infty & \text{otherwise}.
    \end{cases}
    \label{eq:dist}
\end{equation}
The tree of node set $V_s$ can be implemented as an array, then the time of calculating $dist(v,T)$ is reduced from $O(|V_s| + |E_s|)$ to $O(k)$.

When a new edge $(u,v)$ is added into the subgraph $G_s$, the distance between the query node set $T$ and other nodes in the subgraph may change, then the distance tree needs to be updated accordingly. There are two cases in which the distance from $T$ to other nodes will be updated. Without loss of generality, we assume $dist(u, T) < dist(v, T)$. 
One case is that $v$ is not in $V_s$ before, and when $(u,v)$ is added, $dist(v,T)$ should be changed to $dist(u,T) + 1$. Correspondingly, the distance tree needs to add a new node $v$ as a child of node $u$. 
The other is that both $u$ and $v$ are already in $V_s$ and $dist(v,T) - dist(u,T) \geq 2$ before. % $(u,v)$ is added.
When $(u,v)$ is added, $dist(v,T)$ should be changed to $dist(u,T)+1$ and $dist(\cdot,T)$ of other nodes whose shortest paths to $T$ containing $(u,v)$ should also be updated. %after $(u,v)$ is added. 
For this complex situation, the distance tree only needs to change the parent node of $v$ to $u$, and the time complexity of this operation is $O(1)$. 
As the example shown in Fig. \ref{fig:tree}, when a new edge $(2,6)$ is added into the subgraph, the shortest path from node $6$ to the query nodes is changed to $6 \rightarrow 2$.
Thus the parent of node $6$ is changed from node $7$ to node $2$, and we only need to change one element in the array storing the parent node.

\subsection{Community Extraction}
After we sample a subgraph covering the local structure around the query nodes from the graph stream, the second phase is to find the target community $C^*$ containing the query nodes on the sampled subgraph $G_s$. 
Considering that nodes within the same community are closely connected, we adopt  \textit{seed-set expansion} strategy using the query nodes in $T$ as the seed-set.
%and expand the seed-set by adding closely connected nodes to extract the target community $C^*$ on the sampled local structure (or subgraph) $G_s$.
For each node in $G_s$, we evaluate its probability of belonging to the target community by the closeness of connection to the seed-set and add nodes to the seed-set in descending order of the probability by optimizing the community quality.

In the following, we will introduce in detail the probability diffusion method and the proposed \textit{approximate conductance} metric for measuring the community quality, as well as the %complete
algorithm flow.

\subsubsection{Diffusion method}
To measure the closeness of the connection between seed-set $T$ and the remaining nodes in subgraph $G_s$, we employ a lazy random walk for probability diffusion. For each node $u \in T$, we start a lazy random walk from $u$, with $0.5$ probability staying at the current node and $0.5$ probability following an adjacent edge chosen uniformly and randomly.
Repeating this operation several times, we will obtain a probability distribution covering all nodes in subgraph $G_s$, and treat the probability on each node as its closeness to seed-set $T$.

The transition matrix $\mathbf{N_{rw}}$ of the lazy random walk is given by:
\begin{equation}
    \mathbf{N_{rw}} = (\mathbf{I} + \mathbf{D_s}^{-1} \mathbf{A_s})/2,
    \label{eq:transition}
\end{equation}
where $\mathbf{A_s}$ is the adjacent matrix of $G_s$, $\mathbf{D_s}$ is the diagonal matrix of degrees on $G_s$, and $\mathbf{I}$ is an identity matrix of order $|V_s|$.

The initial probability $\mathbf{p} \in [0,1]^{|V_s|}$ is concentrated evenly at the seed-set, defined formally as:
\begin{equation}
    p_i = 
    \begin{cases}
        1 / |T| & \text{if} \quad v_i \in T, \\
        0 & \text{if} \quad v_i \in V_s \setminus T.
    \end{cases}
    \label{eq:prob}
\end{equation}
Since the sampled subgraph $G_s$ only covers the $k$-hop neighborhood of seed-set $T$, we do a $k-$step lazy random walk, and the probability will diffuse to the entire subgraph $G_s$.

\subsubsection{Scoring Function}
\label{sec:scoring}
Suppose the size of the target community (ground truth size) is available. In that case, the community can be obtained by cutting the probability distribution vector $\mathbf{p}$ with the size as a budget. Unfortunately, the target size is not available in many real-world scenarios, especially in naturally streaming networks. To address this issue, we determine the community boundary automatically by optimizing a scoring function.

Modularity~\cite{newman2006modularity}, as a metric evaluating graph partition, is usually adopted by global community detection methods. Conductance~\cite{kannan2004clusterings} is also a community metric highly correlated with the conception of tight internal connections but sparse external connections. Compared to modularity, conductance is usually adopted as the measuring metric for a single community and employed in various local community detection methods as the scoring function~\cite{andersen2006local, li2015uncovering, he2019krylov}. 
The conductance is defined as follows. 
\begin{definition}[Conductance]
Given a graph $G=(V,E)$, for a node set $C$ and $C \subset V$, the conductance of the induced subgraph of $C$ is: 
\begin{equation}
    \label{eq:cond}
    \Phi(C) = \frac{\text{cut}(C, V \setminus C)}{\text{min}\{\text{Vol}(C),\text{Vol}(V \setminus C)\}},
\end{equation}
where $\text{cut}(\cdot, \cdot)$ denotes the number of edges between two node sets and $\text{Vol}(\cdot)$ denotes the total degree of a node set.
\end{definition}

For a large-scale graph, the cost of calculating conductance by (\ref{eq:cond}) is unacceptable. If there is a subgraph covering the local structure around community $C$ (\ie, community $C$ and its adjacent nodes), calculating $\text{Vol}(C)$ on the complete graph can be simplified to calculate locally on the subgraph~\cite{li2015uncovering,he2019krylov}.
When $|C| \ll |V|$, the denominator of $\Phi(C)$ can be simplified to $\text{Vol}(C)$, and this condition is usually satisfied for community detection tasks. 
According to existing practices, the conductance on the entire graph can be simplified to the conductance on the subgraph:
\begin{equation}
    \label{eq:local_cond}
    \Phi(C) = \frac{\text{cut}(C, V_s \setminus C)}{\text{Vol}_\text{s}(C)},
\end{equation}
where $\text{Vol}_s(\cdot)$ denotes the total degree of a node set on the subgraph.

\begin{figure}[htbp]
    \centering
    \includegraphics[width=0.5\linewidth]{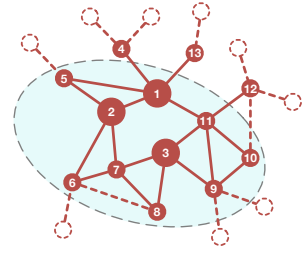}
    \caption{An example of the sampled subgraph, with \{1,2,3\} the query nodes. 
    Sampled nodes/edges are in solid lines, while missed nodes/edges are in dashed lines. 
    }
    \label{fig:cond}
\end{figure}

For graph streams, the entire graph cannot be cached, so the exact conductance is unavailable. The conductance calculation can only be carried out on limited information cached in the graph stream. Unfortunately, compared to a subgraph sampled on a static complete graph, the subgraph sampled in the graph stream is usually incomplete. 
Since edges come in arbitrary order and can only be accessed once, we will miss some useful nodes and edges, which are not connected to nodes that have been sampled into the subgraph before (see Algorithm \ref{alg:sampling}). As illustrated in the example in Fig. \ref{fig:cond}, neither $\text{cut}(C, V_s \setminus C)$ nor $\text{Vol}_\text{s}(C)$ calculated on such sampled subgraph can be guaranteed to be accurate. Consequently, the conductance calculated by (\ref{eq:local_cond}) is inaccurate on the subgraph sampled in graph streams.

Though we cannot cache the complete graph for the graph stream, we observe that some information can still be obtained exactly during the sampling. 
Specifically, we count the times that each node has appeared when processing the graph stream. In the end, we could obtain the degree of each node in $V$, which is stored in the degree array $D$ in Algorithm \ref{alg:sampling}.
And $\text{cut}(C, V \setminus C)$, which represents the number of edges between $C$ and $V \setminus C$, is equivalent to the value that the total number of edges connected with nodes in $C$ minus the number of edges within $C$. %For graph streams,
Hence, we propose an approximate conductance on the sampled subgraph as follows:  
\begin{equation}
    \label{eq:appr_cond}
    \Phi(C) = \frac{\text{Vol}(C) - 2|E_s(C,C)|}{\text{Vol}(C)},
\end{equation}
where $E_s(C,C)$ denotes the set of edges within the node set $C$ on the subgraph $G_s$. 

Compared to existing conductance calculated locally on the sampled subgraph, the approximate conductance is more accurate and suitable for graph streams. There is still a deviation between the approximate conductance and the exact conductance, which is due to the difference of $E_s(C,C)$ and $E(C, C)$,
However, the closer an edge to the query nodes, the less likely it is to be missed during the stream sampling, and the nodes in the community $C$ are usually closely connected to the query nodes, indicating that the approximate conductance can be considered as a qualified deputy for the exact conductance. 

For the example in Fig. \ref{fig:cond}, we regard the nodes within the light green area as the members of $C$. For the exact conductance, $cut(C, V \setminus C) = 8$, $\text{min}\{\text{Vol}(C),\text{Vol}(V \setminus C)\} = \text{Vol}(C) = 40$, and the result is $0.20$. For the existing conductance calculated locally on the sampled subgraph, $cut(C, V_s \setminus C)=3$, $\text{Vol}_\text{s}(C)=33$, and the result is $0.09$. For the approximate conductance, $|E_s(C,C)|=15$, $\text{Vol}(C)=40$, which is derived from $D$, 
% is equal to that for standard conductance on the original graph
and the result is $0.25$. Obviously, the approximate conductance on the sample subgraph is closer to 
the exact conductance on the complete graph.
% the standard conductance on the original graph. 
% \KunHe{changed. check this paragraph}

\subsubsection{Algorithm description}
The probability distribution vector $\mathbf{p}$, which is obtained by a $k-$step lazy random walk, indicates the closeness between nodes in $V_s$ and seed-set $T$, \eg, $p_i > p_j$ indicates $v_i$ is closer to $T$ than $v_j$ and $v_i$ is more likely to be a member of the target community. We sort all nodes in $V_s$ by $\mathbf{p}$ in the descending order, then the community members appear on the front, but we do not know the exact size. To determine the boundary of the community (or the community size), we add the first $i \in \{1, 2, 3, ...\}$ nodes in the sorted $V_s$ into the seed-set as the candidate community, and the one with minimum approximate conductance is the detected community $\hat{C}$. In addition, we introduce an upper bound $b$ for the community size, and we only need to consider candidate communities consisting of the first $i$ ($i \in [1,b]$) nodes in the sorted $V_s$ and $T$.

Details of extracting community on the sampled subgraph $G_s$ are presented in Algorithm \ref{alg:community}.

\renewcommand{\algorithmicrequire}{\textbf{Input:}}
\renewcommand{\algorithmicensure}{\textbf{Output:}}
\begin{algorithm}
\begin{algorithmic}[1]
\caption{Extract community on sampled subgraph}
\label{alg:community}
\REQUIRE Sample subgraph $G_s = (V_s, E_s)$, query nodes $T$, degree array $D$. %$\mathbf{D}$.
\ENSURE Extracted community $\hat{C}$.
\renewcommand{\algorithmicrequire}{\textbf{Parameters:}}
\REQUIRE Number of hops $k$, community upper bound $b$.

\STATE $\mathbf{A_s} \gets \text{adjacency matrix of } G_s$
\STATE $\mathbf{D_s} \gets \text{diagonal degree matrix of } \mathbf{A_s}$
\STATE $\mathbf{N_{rw}} \gets (\mathbf{I} + \mathbf{D_s}^{-1} \mathbf{A_s})/2$
\STATE Initialize $\mathbf{p}$ by (\ref{eq:prob})
\STATE $\mathbf{p} \gets (\mathbf{N_{rw}})^k \mathbf{p}$
\STATE Sort $V_s$ by $\mathbf{p}$ in descending order as $\{ v_1, v_2, ..., v_{|V_s|} \}$.
\FOR{$i \gets 1$ \textbf{to} $b$}
    \STATE $\hat{C}_i \gets \{ v_j | j \leq i \} \cup T$
    \STATE Calculate $\Phi(\hat{C}_i)$ by (\ref{eq:appr_cond}) with $E_s$ and $D$.
\ENDFOR
\STATE $\hat{C} \gets \hat{C}_i$ with the minimum $\Phi(\hat{C}_i)$
\end{algorithmic}
\end{algorithm}

\subsection{Complexity Analysis}
\label{sec:complexity}
Here we analyze the computational time and space complexity of the proposed algorithm. %And we consider the task of revealing the target community for a single query node set $T$ given the graph stream $\mathcal{S}$ of $G=(V,E)$.

\subsubsection{Time complexity}
\label{sec:time_complexity}
Our method mainly includes two phases: sampling subgraph in graph stream and extracting community in the sampled subgraph. According to Algorithm \ref{alg:sampling}, the computational time of each coming edge is bounded by the time of calculating $dist(\cdot, T)$, and the time complexity of $dist(\cdot, T)$ is $O(k)$ in the worst case (as discussed in Section \ref{sec:sampling}). 
In addition, we need to prune the sampled subgraph after processing every $pc$ edges, and calculate $dist(\cdot, T)$ for $O(|V_s|\log|V_s|)$ times for each pruning.
So we can sample the subgraph in $O(|E|k+(|E|/pc)(|V_s|\log|V_s|)k)$ for the query nodes. 

According to Algorithm \ref{alg:community}, the computational time is bounded by line $5$ and line $7-9$. In line $5$, we need to multiply the transition matrix $\mathbf{N_{rw}}\in\mathbb{R}^{|V_s|\times|V_s|}$ and probability vector $\mathbf{p}\in[0,1]^{|V_s|}$ for $k$ times, whose time complexity is $O(k|V_s|^2)$. Line $7-9$ calculates the approximate conductance for $b$ times and the time of each calculation is bounded by $O(b^2)$, so the total is $O(b^3)$.

In summary, the time complexity of our method is $O(k|E|(1 + (|V_s| \log |V_s|)/pc) + k|V_s|^2 + b^3)$. As $k$ is a small integer and $|V_s| \ll |E|, b \ll |E|$, the time complexity is linear to the stream size $|E|$.

\subsubsection{Space complexity}
As shown in Algorithm \ref{alg:sampling}, when sampling the subgraph, the degree array $D$ requires $O(V)$ space, and the sampled subgraph requires $O(|V_s|+|E_s|)$ space. During the community extraction on the sampled subgraph, as shown in Algorithm \ref{alg:community}, the edge set $E_s$ is replaced by the adjacency matrix $\mathbf{A_s}$ with $O(|V_s|^2)$, probability vector $\mathbf{p}$ and storing $b$ approximate conductance values will occupy $O(|V_s|)$ and $O(b)$ respectively. 

In summary, the space complexity of our method is $O(|V|)$, which satisfies the work memory limitation of graph streams~\cite{mcgregor2014graph}.

\section{Experiments}
% consider this section consist of two part: one is comparison with CoEuS on accuracy (F1-score), time (running time), space (main memory); the other is comparison with typical (or other word) method on accuracy (F1-score), space (main memory). The comparison with traditional method is call back the limitation of traditional method mentioned in INTRODUCTION.
% Above discussion may not work, this paper is focus on situation of graph stream, and the traditional method cannot work in graph stream. So we only need compare to the method on graph stream, CoEuS, and rest focus on present our insight of community on graph stream by case study or other love experiments.

% show the time comparison with CoEuS in stacked bar with log-scale (below is the time of stream processing, above is the time of community detection). Ours is faster than CoEuS, especially on stream processing, which is importance for effectiveness of online algorithm and it need to match the arrival speed of the graph flow. 
In this section, we conduct comprehensive experiments and focus on answering the following questions:
\begin{itemize}
    \item How effective is the approximate conductance?
    \item How does our method compare to the existing streaming local community detection method on both effectiveness and efficiency?
    \item Can our method in graph streams has \textit{similar} performance to the non-streaming local community detection methods that run on the static complete graph? 
    \item How about the scalability and preference of our method?
    \item How to choose appropriate hyper-parameter settings? 
\end{itemize}

\subsection{Experimental Setup}
\subsubsection{Datasets}
To verify the effectiveness and efficiency of our algorithm in real-world scenarios, we adopt five real-world networks with ground truth community membership from the Stanford Network Analysis Project (abbreviated as SNAP)\footnote{https://snap.stanford.edu/data/\#communities}. These networks cover various categories of network applications, including product networks (Amazon), collaboration networks (DBLP), and social networks (Youtube, LiveJournal, and Orkut). 
To ensure a fair comparison, we follow the pre-processing procedure of existing local community detection methods~\cite{li2015uncovering,liakos2020rapid}, only adopting the top 5000 ground truth communities with the highest quality according to \cite{yang2015defining}, and meanwhile discard communities with size less than 20. 
The statistical information of these networks (number of nodes/edges/ground truth communities) is summarized in Table \ref{tab:datasets}, and all networks are considered undirected and unweighted. 

\begin{table}
    \caption{Statistics of Real-world Networks.}
    \label{tab:datasets}
    \centering
    \begin{tabular}{c r r r}
    \toprule
    Network & \# Nodes & \# Edges & \# Communities\\
    \midrule
    Amazon & 334,863 & 925,872 & 936\\
    DBLP & 317,080 & 1,049,866 & 283\\
    Youtube & 1,134,890 & 2,987,624 & 652\\
    LiveJournal & 3,997,962 & 34,681,189 & 2,159\\
    Orkut & 3,072,441 & 117,185,083 & 4,530\\
    \bottomrule
    \end{tabular}
\end{table}

Unless otherwise specified, for all the experiments in this paper, we randomly select 500 ground truth communities for each network as test cases. For each test case, we randomly select three nodes from the community as the query nodes. If the total number of ground truth communities in a particular network is less than 500, we select all ground truth communities in the network as its test cases.

% \begin{table}
%     \caption{Statistics of Real-world Networks.}
%     \label{tab:datasets}
%     \begin{tabular}{c l r r}
%     \toprule
%     Category &Network & \# Nodes & \# Edges\\
%     \midrule
%     Product & Amazon & 334,863 & 925,872 \\
%     Collaboration & DBLP & 317,080 & 1,049,866\\
%     \multirow{3}*{Social} & Youtube & 1,134,890 & 2,987,624\\
%     ~ & LiveJournal & 3,997,962 & 34,681,189\\
%     ~ & Orkut & 3,072,441 & 117,185,083\\
%     \bottomrule
%     \end{tabular}
% \end{table}

\subsubsection{Baselines}
To give a well-round comparison, we consider two types of local community detection methods. 
\begin{itemize}
    \item \textbf{Streaming method} work in graph streams. To the best of our knowledge, there is only one existing streaming local community detection method called CoEuS~\cite{liakos2020rapid}, which we select as the baseline.
    \item \textbf{Non-streaming methods} run on \textit{static} and \textit{complete} graphs. 
    There are two state-of-the-art local community detection algorithms, LEMON~\cite{li2015uncovering}
    and LOSP~\cite{he2019krylov}, which we choose as the reference. 
\end{itemize}

\subsubsection{Settings}
Our SCDAC method is implemented in Python 3 with numpy library%\footnote{\url{https://drive.google.com/file/d/1n0gCz9q_6jcgciGxwAh2a3eN-VnpdL66/view?usp=sharing}}. 
There are four hyper-parameters in our method, including the number of hops $k=4$, pruning cycle $pc = 100,000$, pruning size $ps = 3,000$ and the upper bound of community size $b = 500$. For all baselines, we keep the default parameter settings across all experiments. Moreover, as LOSP considers small-scale communities while we follow the dataset settings of LEMON~\cite{li2015uncovering} and CoEuS~\cite{liakos2020rapid} (which only consider communities with size greater than 20), we change the minimal communities size of LOSP from 3 to 20.
Following the existing local community detection works~\cite{li2015uncovering,he2015detecting,he2019krylov,liakos2020rapid}, we adopt the F1-score between the detected community and the target community to evaluate the effectiveness of the methods.

The network data file downloaded from SNAP~\footnotemark[1] is organized by edges of the network in lines. And the graph stream is generated by the edges read line by line from the network data file. To ensure the validity of the experimental results, each experiment in this work has been repeated many times. Each time we randomly shuffle the order of edges in the graph stream and randomly select the test cases and the query nodes again. And the experimental results are reported in the mean $\pm$ standard error of the average of the test cases.

All the experiments in this section are conducted on a machine with 2 Intel Xeon CPUs at 2.3GHZ and 256GB main memory.

\subsection{Evaluation on Approximate Conductance}
\begin{figure}
    \centering
    \includegraphics[width=\linewidth]{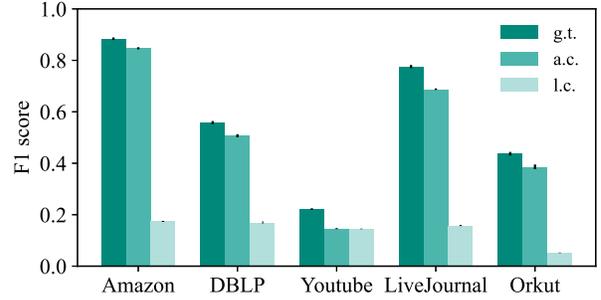}
    \caption{Comparison of F1-score for different scoring functions.}%with different scoring functions.}
    \label{fig:f1_cond}
\end{figure}

% comparison between community extraction with approximate conductance and local conductance.
To verify the effectiveness of approximate conductance, we compare the performance of community extraction with approximate conductance and existing local conductance as scoring functions and directly cut the probability distribution vector $\mathbf{p}$ with the ground truth size. Fig. \ref{fig:f1_cond} shows the comparison of F1-score on real-world networks, where ``g.t.'', ``a.c.'' and ``l.c.'' denote the results of SCDAC to detect communities by cutting the probability distribution vector with the ground truth size, adopting approximate conductance (in Equation \ref{eq:appr_cond}) and adopting locally calculated conductance (in Equation \ref{eq:local_cond}) as the scoring function, respectively.

Obviously, the F1-score of adopting the local conductance as the scoring function is pretty low across all real-world networks, which implies that the local conductance, defined on the sampled subgraph from the streaming data, may not accurately measure the community quality, neither is it suitable as the scoring function for seed-set expansion methods for graph streams (this also verifies our discussion in Section \ref{sec:scoring}).
In contrast, the F1-score of adopting approximate conductance as the scoring function is much higher than that of the local conductance, and it is close to the F1-score of extracting communities by cutting $\mathbf{p}$ with ground truth sizes. Specifically, for all real-world networks, the gap between them is no more than 0.09, and the average value of the gap is only 0.06. The experimental result indicates that our proposed ``approximate conductance'' can be a good measure for community quality on the subgraph sampled from the graph streams. It also suggests that our community extraction method can be applied to reveal the target community for the query nodes in graph streams of real-world networks without the supervision of ground truth size.

\subsection{Comparison with Streaming Method}
We conduct experiments on five real-world networks and comprehensively compare our method with a streaming local community detection method called CoEuS~\cite{liakos2020rapid} without the supervision of ground truth size for both effectiveness and efficiency.

\subsubsection{Effectiveness}
\begin{figure}
    \centering
    \includegraphics[width=\linewidth]{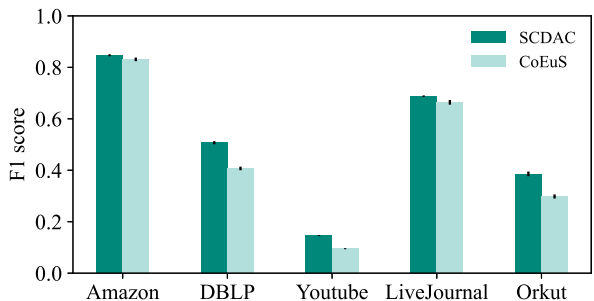}
    \caption{Comparison of F1-score with the streaming local community detection methods in graph streams.}
    \label{fig:f1_stream}
\end{figure}

% F1-score comparison between SCDAC and CoEuS.
We report the F1-score as the metric of effectiveness, and the comparison results are shown in Fig. \ref{fig:f1_stream}. It can be observed that our SCDAC method outperforms CoEuS across all networks. Especially on DBLP network, the F1-score of SCDAC exceeds that of CoEuS by 0.10, and on Youtube and Orkut networks, the F1-score of CoEuS is only about three-quarters of that of SCDAC. Compared to CoEuS, SCDAC can reveal the target local community in graph streams more accurately.

\subsubsection{Efficiency}
% running time comparison between SCDAC and CoEuS
We report the average running time of detecting a single community (for a single test case) as the metric of efficiency. To ensure a fair comparison, SCDAC and CoEuS are all implemented in Python 3, and they both uncover the community for all test cases at once on each network.

Table \ref{tab:time_streaming} illustrates the comparison of running time. It shows that SCDAC is considerably faster than CoEuS on most networks. Specifically, on Amazon network, SCDAC is twice as fast as CoEuS; on LiveJournal and Orkut networks, SCDAC only takes two-thirds and three-quarters of the corresponding running time of CoEuS, respectively. And CoEuS is only faster than SCDAC on the Youtube network; this is because that the Youtube network is sparsely connected (the average degree is only 2, refer to Table \ref{tab:datasets}) and the expansion of CoEuS, which is in a greedy way only needs to consider a limited number of edges.

% \begin{table}
%     \caption{Comparison of running time (in second) for streaming local community detection methods.}
%     \centering
%     \begin{tabular}{c c c c c c}
%     \toprule
%     Method & Amazon & DBLP & Youtube & LiveJournal & Orkut \\
%     \midrule
%     % SCDAC & 1.10 & 2.05 & 13.25 & 38.23 & 204.13 \\
%     % CoEuS & 2.56 & 3.00 & 19.45 & 78.52 & 276.76 \\
%     \bottomrule
%     \end{tabular}
%     \label{tab:time_streaming}
% \end{table}
% SCDAC & 0.95$\pm$0.00 & 2.10$\pm$0.03 & 15.79$\pm$0.26 & 62.93$\pm$3.57 & 247.48$\pm$25.31 \\
% CoEuS & 2.02$\pm$0.12 & 2.46$\pm$0.02 & 7.50$\pm$0.27 & 95.25$\pm$14.56 & 318.19$\pm$19.98 \\
\begin{table}
    \caption{Comparison of running time (in second) for streaming local community detection methods.}
    \centering
    \begin{tabular}{c c c}
        \toprule
        Network & SCDAC & CoEuS \\
        \midrule
        Amazon & 0.95 $\pm$ 0.00 & 2.02 $\pm$ 0.12  \\
        DBLP & 2.10 $\pm$ 0.03  & 2.46 $\pm$ 0.02  \\
        Youtube & 15.79 $\pm$ 0.26  & 7.50 $\pm$ 0.27  \\
        LiveJournal & 62.93 $\pm$ 3.57  & 95.25 $\pm$ 14.56  \\
        Orkut & 247.48 $\pm$ 25.31  & 318.19 $\pm$ 19.98  \\
        %Amazon & 1.10 secs. & 2.56 secs. \\
        %DBLP & 2.05 secs. & 3.00 secs. \\
        %Youtube & 13.25 secs. & 19.45 secs. \\
        %LiveJournal & 38.23 secs. & 78.52 secs. \\
        %Orkut & 204.13 secs. & 276.76 secs. \\        
        \bottomrule
    \end{tabular}
    \label{tab:time_streaming}
\end{table}

In summary, SCDAC \textit{remarkably} outperforms the existing streaming local community detection method in terms of effectiveness and efficiency. Experimental results also imply that SCDAC can effectively and quickly solve the streaming local community detection problem for massive real-world networks. 

\subsection{Comparison with Non-streaming Method}
\begin{figure}
    \centering
    \includegraphics[width=\linewidth]{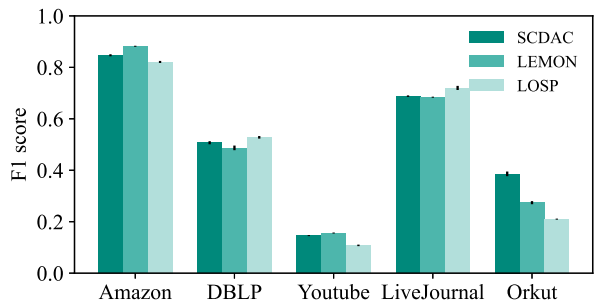}
    \caption{Comparison of F1-score with the non-streaming local community detection methods. %which runs on static, complete graphs.
    }
    \label{fig:f1_typical}
\end{figure}

% F1-score comparison between SCDAC, LEMON and LOSP.
% experimental settings
To further test the effectiveness of our method, we compare the performance of SCDAC and two state-of-the-art non-streaming local community detection methods called LEMON~\cite{li2015uncovering} and LOSP~\cite{he2019krylov} on the same networks. Note that SCDAC works in graph streams while LEMON and LOSP run on the static and complete graphs.
Due to the strict restrictions in graph streams, SCDAC faces tougher challenges than the methods under the non-streaming scenarios. Fig. \ref{fig:f1_typical} illustrates the comparison of F1-score between SCDAC and the reference methods.

% describe the experimental results and notice the space overhead.
As we can see in Fig. \ref{fig:f1_typical}, SCDAC, LEMON, and LOSP show similar performance on Amazon, DBLP, Youtube, and LiveJournal networks, but on Orkut network, SCDAC is much superior to LEMON and LOSP. To be precise, on Amazon and Youtube networks, the F1-score of SCDAC is 0.023 lower than that of LEMON on average. While on DBLP, LiveJournal, and Orkut networks, SCDAC outperforms LEMON on the F1-score by an average of 0.045.
As for LOSP, it performs better on DBLP and LiveJournal networks and performs inferior to SCDAC on Amazon, Youtube, and Orkut networks. Especially on the Orkut network, SCDAC outperforms LOSP by 0.176 on the F1-score.

These experimental results have verified that SCDAC, working in graph streams with access constraints and pretty low memory, can still achieve \textit{similar} performance to the state-of-the-art non-streaming local community detection method. It also indicates that SCDAC can accurately uncover the target community for the query nodes even in graph streams and can be competent for the task of local community detection on \textit{massive} and \textit{naturally streaming} networks.

\subsection{Scalability Testing}
We further analyze the scalability of the proposed method. To this end, We generate several synthetic datasets and analyze the performance of our method in different configurations. For synthetic datasets, we employ the LFR benchmark networks~\cite{lancichinetti2008benchmark}, which simulates some properties of real-world networks, \eg the heterogeneity of node degrees and community size distribution.

\subsubsection{Scalability on network scale and density}
To investigate the scalability of SCDAC on the network scale and density, we compare the running time of SCDAC on networks with different numbers of nodes or average degrees. For the corresponding two groups of networks, we use the same basic parameter settings: the number of nodes of 1,000,000, the average degree of 20, the maximum degree of 100, the mixing parameter of 0.1, and the community size is within $[20, 500]$. Specifically, for generating the networks with various network scales, we vary the number of nodes from 200,000 to 2,000,000, and other parameters follow the basic settings. For generating the networks with various average degrees, we vary the average degree from 12 to 30, and other parameters follow the basic settings.

\begin{figure}
    \centering
    \subfigure[Number of nodes]{
        \includegraphics[width=0.46\linewidth]{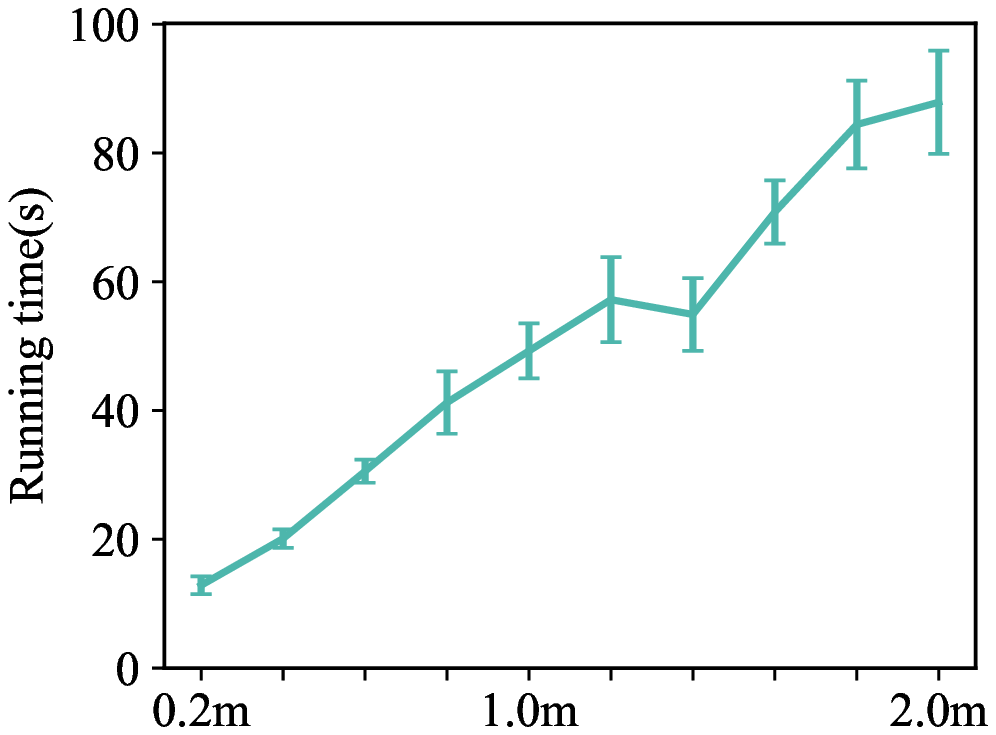}
    }
    \subfigure[Average degree]{
        \includegraphics[width=0.46\linewidth]{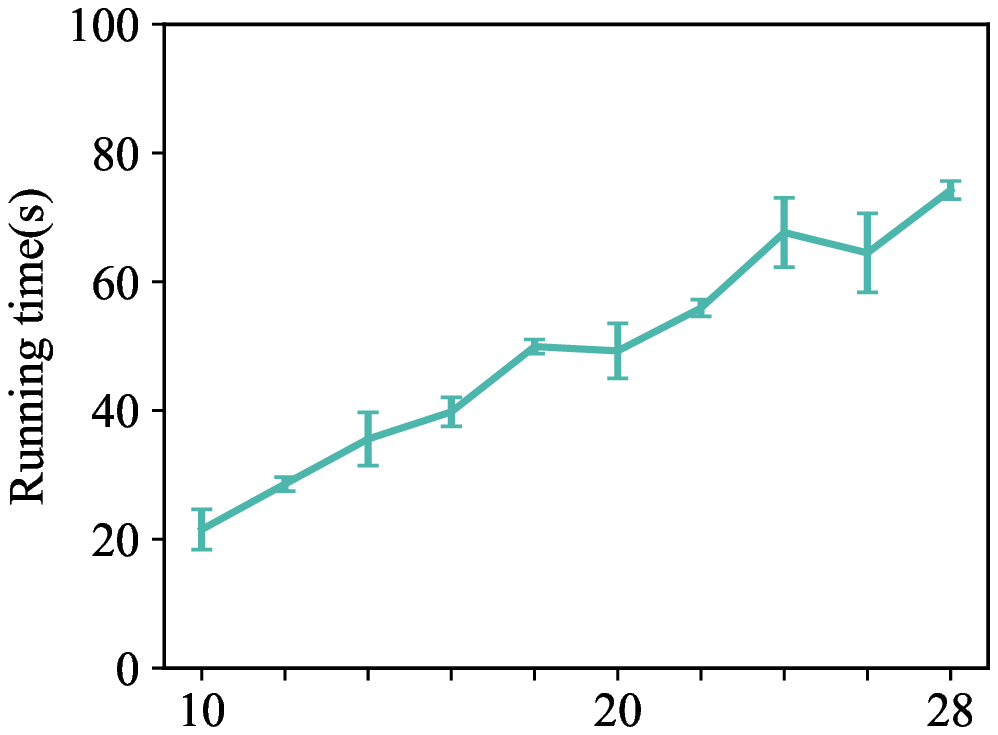}
    }
    \caption{Running time of SCDAC on networks with various scales or densities.}
    \label{fig:scalability_networks}
\end{figure}

Fig.~\ref{fig:scalability_networks} illustrates the average running time of detecting a single community (for a single test case) on networks with different numbers of nodes or different average degrees. As can be seen, with the increase of the network scale, the time cost grows almost linearly, and the running time of SCDAC is also linear to the average degree. 
The two results are consistent with our time complexity analysis in Section \ref{sec:time_complexity}.
%that the running time of SCDAC is linear to the number of network edges. 
In summary, SCDAC exhibits good scalability from the perspective of network scale and network density.

\subsubsection{Scalability on community size and density}
To investigate the preference of SCDAC for different communities, we compare the F1-scores for detecting communities in different sizes or densities. The two experiments are carried out on the same set of networks, but the communities with different sizes or densities are selected as the test cases correspondingly. For the network, we still use the previous set of parameter settings, that is, the number of nodes is 1,000,000, the average degree is 20, the maximum degree is 100, the mixing parameter is 0.1, and the community size is within $[20, 500]$.

\begin{figure}
    \centering
    \subfigure[Number of nodes]{
        \includegraphics[width=0.46\linewidth]{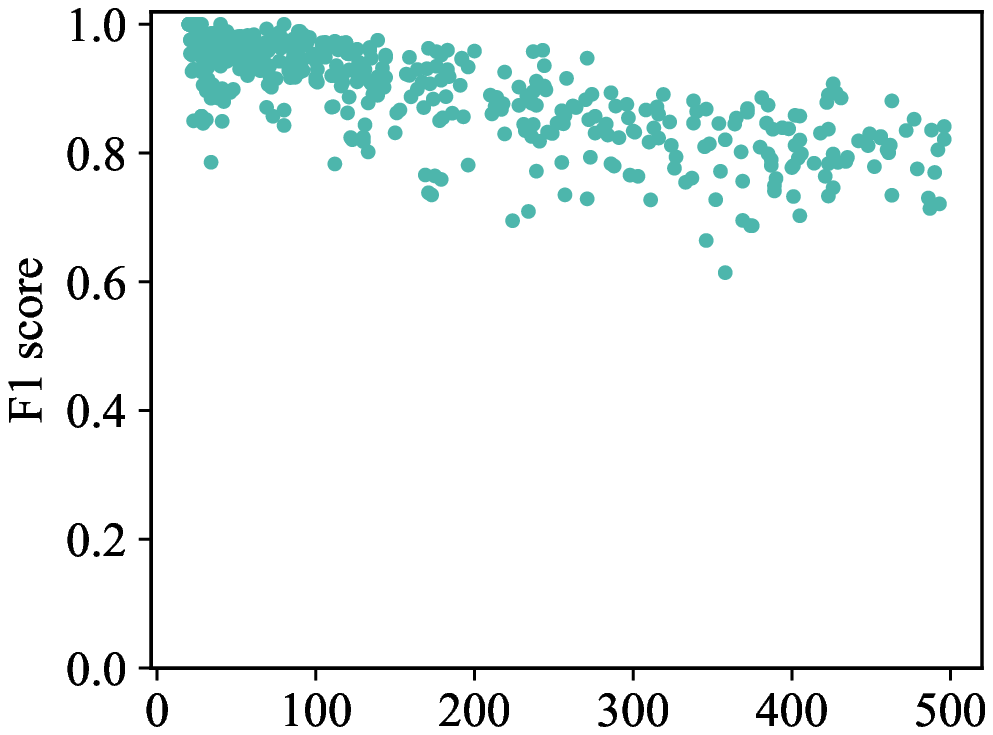}
    }
    \subfigure[Density]{
        \includegraphics[width=0.46\linewidth]{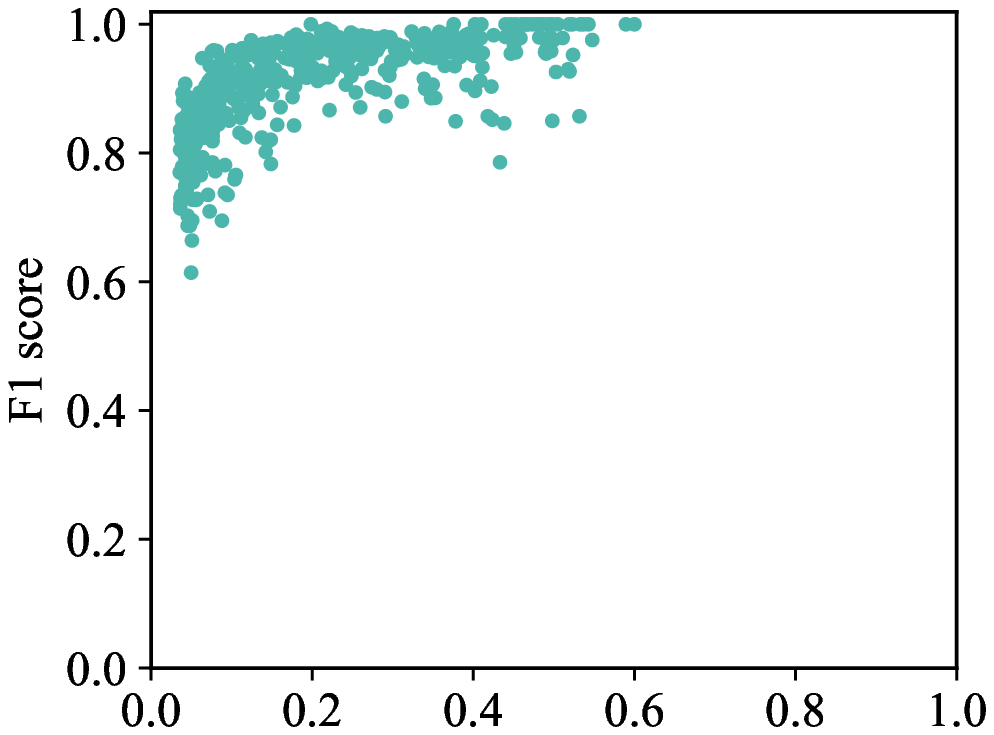}
    }
    \caption{Performance of SCDAC on communities with various sizes or densities.}
    \label{fig:scalability_communities}
\end{figure}

Fig.~\ref{fig:scalability_communities} (a) demonstrates the distribution of F1-scores with respect to the number of nodes in the target community. One can observe that as the number of nodes grows, the F1-score shows a slightly downward trend. For communities with less than 100 nodes, the F1-score of SCDAC is around 0.9, while for communities with 400 to 500 nodes, the F1-score of SCDAC decays to about 0.8. 
The results are acceptable as it is usually harder for algorithms to detect large communities accurately, as larger communities will be affected more by the neighborhood.
%The reason for this phenomenon is that SCDAC detects communities on the fixed-scale sampled subgraphs. Then the larger the community, the harder it is for the sampled subgraph to cover all nodes in the community, thereby affecting the community detection results.

The distribution of F1-scores with respect to the density of communities is shown in Fig.~\ref{fig:scalability_communities} (b). SCDAC performs better on denser communities. SCDAC samples subgraphs in the graph stream based on the distance to the query node(s). Then, the denser the connection with the query node(s), the more likely it could be sampled into the subgraph and is also easier to be detected. And density is a measure of closeness of the connections between members of a community. Hence, SCDAC is more accurate in detecting dense communities.%, and it will not appear in the detected community. 

\subsection{Ablation Study on Parameters}
We carry out an ablation study on the four hyper-parameters of our method, as shown in Algorithm \ref{alg:sampling} and Algorithm \ref{alg:community}.

\subsubsection{Number of hops $k$}
The number of hops $k$ directly determines the neighborhood range of the query nodes we consider, but it does not imply that a larger $k$ is better. When $k$ becomes larger, the computational cost will also increase. As long as $k$ is equal to the diameter of the community (i.e. maximum pairwise distance among nodes in the community), the sampled subgraph can cover the entire community. And as discussed in \cite{conte2018d2k}, the diameter of communities is usually small. Thus, a small positive integer is suitable.

\begin{figure}
    \centering
    \subfigure[Number of hops $k$]{
        \includegraphics[width=0.46\linewidth]{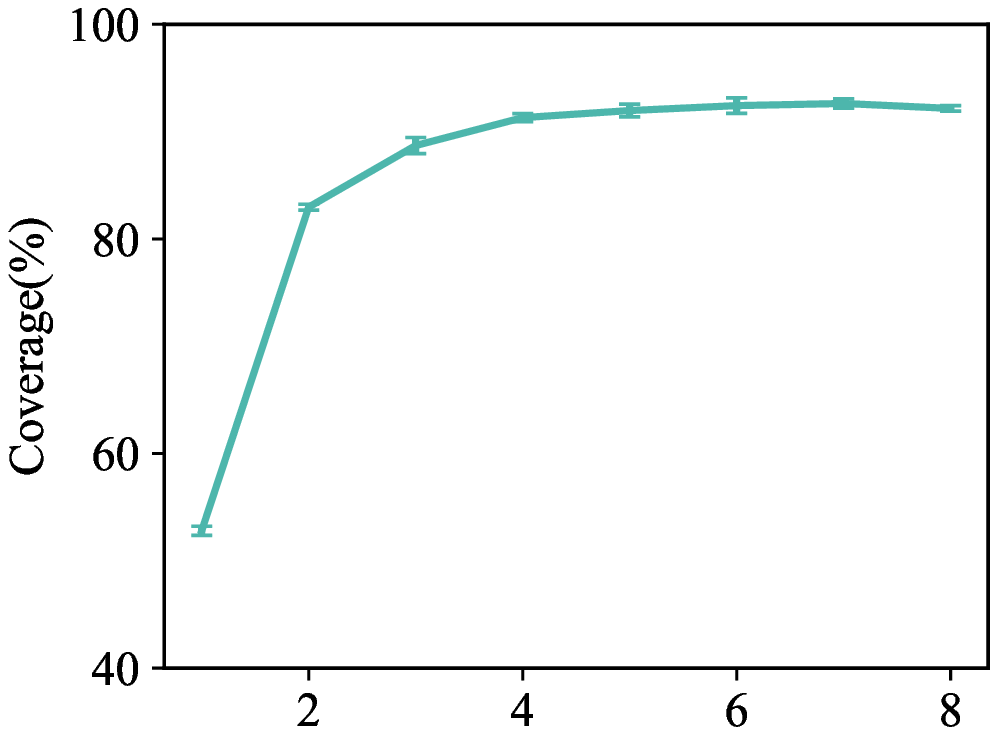}
    }
    \subfigure[Pruning cycle $pc$]{
        \includegraphics[width=0.46\linewidth]{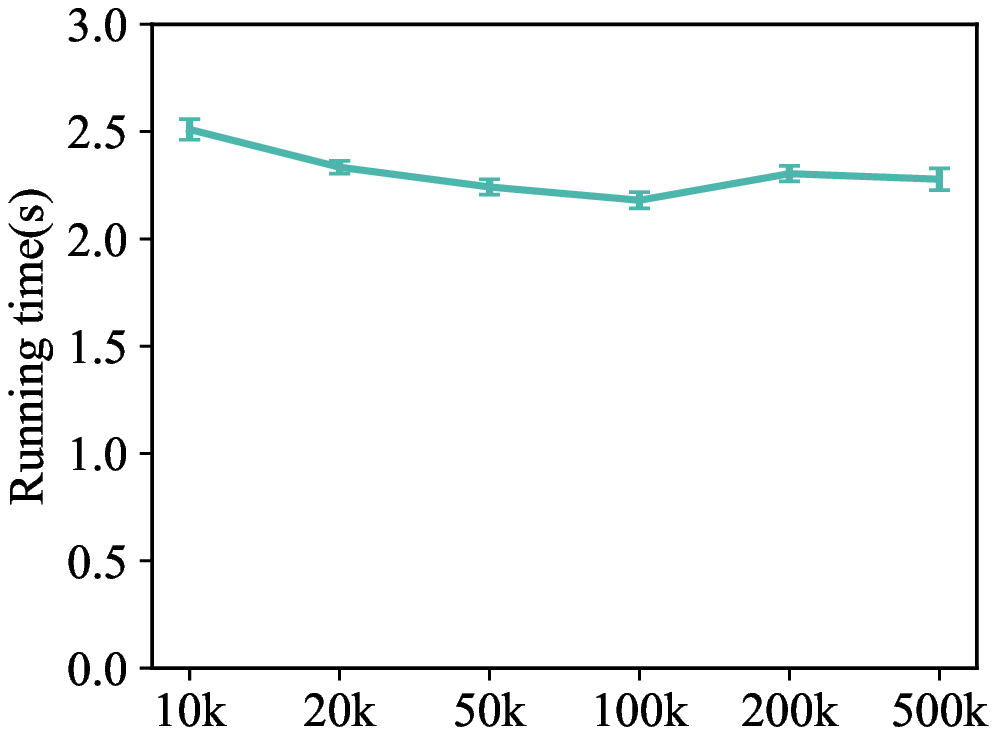}
    }
    \caption{Parameter study on SCDAC.}
    \label{fig:paramters}
\end{figure}
% \Yang{This figure need to be replaced by that with two line and two y-axis(one for time and the other for F1-score).}

To choose an appropriate value for $k$, we vary $k$ from 1 to 8 and evaluate performance by the coverage of sampled subgraph to the target community. Fig. \ref{fig:paramters} (a) demonstrates the results on Amazon network. We observe that with the increase of $k$, when $k \leq 4$, the coverage increases rapidly, but when $k > 4$, the rate of coverage growth starts to be flattened. Considering both the performance and computational cost, $k=4$ is an appropriate setting.

\subsubsection{Pruning cycle $pc$ and pruning size $ps$}
The pruning size $ps$ controls the size of the sampled subgraph for community extraction, and we set $ps = 3,000$, which is an appropriate size that has been verified and applied in prior work~\cite{he2015detecting,he2019krylov}. When SCDAC is applied to detect large-scale communities with thousands of members, it is advisable to turn up the ps so that the sampled subgraph can cover as many members of the target community as possible. Certainly, as $ps$ increases, the time and space overhead of subgraph sampling will increase.

The pruning cycle $pc$ controls the frequency of the pruning process. It is a trade-off for accuracy and efficiency. As $ps$ increases, the sampled subgraph can be more accurately represent the local structure around query nodes. However, as $pc$ increases, the running time of SCDAC is not monotonously increasing. When $pc$ increases, the number of prunings decreases. At the same time, the subgraph grows larger before pruning which will increase the computational cost of the pruning process. We conduct experiment on DBLP network by varying $pc$ from 10,000 to 500,000, and the results are shown in Fig. \ref{fig:paramters} (b). We can observe that SCDAC exhibits the lowest running time when the pruning cycle is in the range of 50,000 to 100,000, and the running time is slightly slower when $pc = 100,000$.

Moreover, $pc$ also affects the memory overhead and accuracy. When SCDAC is running on a machine with very limited memory, it is advisable to turn down $pc$ to reduce the memory cost. And when memory is relatively sufficient, you can turn up $pc$ to obtain more accurate results.

\subsubsection{Upper bound of community size $b$}
Leskovec \etal~\cite{leskovec2008statistical} observe that communities in real-world networks with high quality are small and usually contain no more than 100 nodes. Thus $b$ needs not be too large, or else it will cause a lot of unnecessary computational overhead (as line $7-10$ in Algorithm \ref{alg:community}).

We also statistically analyze the distribution of community sizes in real-world networks, as shown in Table \ref{tab:distribution}. There are at least $97\%$ of communities whose sizes do not exceed 100, except for the DBLP network ($92\%$). And there are at least $97\%$ of communities whose sizes do not exceed 500 in all networks. Thus we set the upper bound of community size to 500.

\begin{table}
    \centering
    \caption{The distribution of ground truth community sizes.}
    \label{tab:distribution}
    \begin{tabular}{c c c c c}
        \toprule
        Network &$[0,100)$ &$[100,500)$ &$[500,1000)$ &$>1000$\\
        \midrule
        Amazon &97.77\% &1.56\% &0.28\% &0.39\% \\
        DBLP &92.88\% &4.68\% &1.33\% &1.11\% \\
        Youtube &99.17\% &0.74\% &0.07\% &0.02\% \\
        LiveJournal &98.59\% &1.19\% &0.14\% &0.08\% \\
        Orkut &99.41\% &0.47\% &0.06\% &0.06\% \\
        \bottomrule
    \end{tabular}
\end{table}

\section{Conclusion}
We propose a novel online, single-pass %paradigm 
method for addressing the local community detection in graph streams. 
The key issue is how could we sample edges under the memory limit and access constraint to build a local structure such that edges close to the query nodes are more likely to be added in the sampled subgraph. 
We design a distance tree that could be re-organized quickly when a coming edge is judged to be added to the subgraph, and we do subgraph pruning periodically to have a good trade-off for accuracy and efficiency on the sampling. 
As edges are coming in arbitrary order, another challenge is how to design a metric that could measure the community quality as accurate as possible. We count the node degree while sampling, and introduce a new metric called ``approximate conductance'' to determine the community boundary on the sampled subgraph. Extensive experiments on real-world datasets show that our method outperforms existing streaming local community detection method with higher accuracy and shorter running time, and even achieves similar performance comparing to the state-of-the-art non-streaming local community detection methods that work on static and complete graphs. We also show that the proposed method is scalable on the network scale and density, and is robust on the detection accuracy for various community sizes and densities. Further parameter studies show that the proposed method is not sensitive to the parameters and exhibits good performance on a suitable range for the parameters. 

Streaming local community detection is a new problem for the massive graph data that grows rapidly, and there are many directions to be explored. Considering the phenomenon that some nodes belong to multiple communities simultaneously, in future work, we will extend our method to uncover multiple communities for query nodes in graph streams.

% or


% \appendices


\ifCLASSOPTIONcompsoc
  \section*{Acknowledgments}
\else
  \section*{Acknowledgment}
\fi

This work is supported by National Natural Science Foundation of China (61772219, U1836204).

\ifCLASSOPTIONcaptionsoff
  \newpage
\fi



%\bibliographystyle{IEEEtran}
\bibliographystyle{IEEEtran}
\bibliography{main}

\vfill


\end{document}